\documentclass[%
superscriptaddress,
reprint,
amsmath,amssymb,
aps,
prr,
]{revtex4-2}

\usepackage[english]{babel}

\usepackage{graphicx}
\usepackage[percent]{overpic}
\usepackage{dcolumn}
\usepackage{bm}
\usepackage{amsmath}
\usepackage{amssymb}

\usepackage{import}
\usepackage{pgf}

\usepackage{hyperref}
\usepackage{color}
\definecolor{rltred}{rgb}{0.75,0,0}
\definecolor{rltgreen}{rgb}{0,0.6,0}
\definecolor{rltblue}{rgb}{0.3,0.3,1}

\usepackage{placeins}

\hypersetup{colorlinks,%
	hypertexnames=true,%
	linkcolor=rltred,%
	citecolor=rltgreen,%
	linktocpage=true}

\newcommand{\tdtwordm}{\textsc{TD2RDM}}
\newcommand{\prdm}[1]{{$#1$\textsc{RDM}}}
\newcommand{\phrdm}[1]{{$#1$\textsc{HRDM}}}
\newcommand{\nrepresentability}{$N$-representability}
\newcommand{\specialterm}[1]{\textit{#1}}
\newcommand{\pdv}[1]{\partial_{#1}}
\DeclareMathOperator{\Tr}{Tr}
\DeclareMathOperator{\id}{id}
\DeclareMathOperator{\Span}{span}

\begin{document}
\title{Projective purification of correlated reduced density matrices}

\author{Elias Pescoller}
\email{elias.pescoller@tuwien.ac.at}
\affiliation{Institute for Theoretical Physics, Vienna University of Technology,
	Wiedner Hauptstra\ss e 8-10/136, 1040 Vienna, Austria, EU}
 
\author{Marie Eder}
\affiliation{Institute for Theoretical Physics, Vienna University of Technology, Wiedner Hauptstra\ss e 8-10/136, 1040 Vienna, Austria, EU}

\author{Iva B\v rezinov\'a}
\email{iva.brezinova@tuwien.ac.at}
\affiliation{Institute for Theoretical Physics, Vienna University of Technology,
	Wiedner Hauptstra\ss e 8-10/136, 1040 Vienna, Austria, EU}

\date{\today}

\begin{abstract}
In the search for accurate approximate solutions of the many-body Schr\"odinger equation, reduced density matrices play an important role, as they allow to formulate approximate methods with polynomial scaling in the number of particles. However, these methods frequently encounter the issue of $N$-representability, whereby in self-consistent applications of the methods, the reduced density matrices become unphysical. A number of algorithms have been proposed in the past to restore a given set of $N$-representability conditions once the reduced density matrices become defective. However, these purification algorithms have either ignored symmetries of the Hamiltonian related to conserved quantities, or have not incorporated them in an efficient way, thereby modifying the reduced density matrix to a greater extent than is necessary. In this paper, we present an algorithm capable of efficiently performing all of the following tasks in the least invasive manner: restoring a given set of $N$-representability conditions, maintaining contraction consistency between successive orders of reduced density matrices, and preserving all conserved quantities. We demonstrate the superiority of the present purification algorithm over previous ones in the context of the time-dependent two-particle reduced density matrix method applied to the quench dynamics of the Fermi-Hubbard model.
\end{abstract}

\maketitle

\section{Introduction}\label{sec:intro}
One of the greatest challenges in theoretical physics to date is the development of accurate approximate solutions to the Schr\"odinger equation for electronic systems. The ability to accurately describe the properties of systems as diverse as multi-electron atoms, molecules, nano-systems, and solids in and out of equilibrium has the potential to facilitate significant future discoveries and technological applications. Consequently, considerable effort has been devoted to the development of methods for obtaining approximate solutions to the Schr\"odinger equation at various levels of complexity.\\
A large class of these approximate methods avoids the exponential scaling of the Schr\"odinger equation by using a reduced object instead of the many-body wavefunction $|\Psi\rangle$. The reduced object is obtained by tracing out most of the degrees of freedom of $|\Psi\rangle$. A prominent example is density-functional theory (DFT) \cite{parr_density-functional_1989} and its time-dependent extension (TDDFT) \cite{ullrich_time-dependent_2011}, which uses the smallest non-trivial object derivable from $|\Psi\rangle$, i.~e.~the particle density. Within the Kohn-Sham ansatz, (TD)DFT scales linearly with the number of particles and is thus the only method to date applicable to bulk solid state systems. Other examples of reduced objects are reduced density matrices (RDMs) \cite{coleman_reduced_2000,mazziotti_advances_2007,bonitz_quantum_2016}, most importantly the one-particle RDM (1RDM) and the two-particle RDM (2RDM), or one- and two-body propagators within (non-equilibrium) Green's function methods \cite{stefanucci_nonequilibrium_2013}. The use of a reduced object instead of the wavefunction typically leads to a polynomial scaling of the methods with particle number. Furthermore, for the majority of physical observables of interest, a reduced object is sufficient to determine the expectation values of these observables, and the knowledge of the full wavefunction is not required, or even not desirable.\\ 
However, eliminating the wavefunction comes at a price. On the one hand, the equations to be solved for the reduced objects are either not known, as in (TD)DFT, or have to be approximated, as in (non-equilibrium) Green's function methods \cite{stefanucci_nonequilibrium_2013} and in approximate closures of the Bogoliubov-Born-Green-Kirkwood (BBGKY) hierarchy for RDMs \cite{bonitz_quantum_2016}. Furthermore, the absence of a reference to a many-body wavefunction inherent to these methods places the issue of \nrepresentability{} \cite{coulson_present_1960, coleman_structure_1963} at the forefront. \nrepresentability{} refers to the question of what are the necessary and sufficient properties of a reduced object such that it can be obtained from a pure wavefunction (pure state \nrepresentability{}) or a many-body density matrix (ensemble \nrepresentability{}) upon tracing out degrees of freedom. While the ensemble \nrepresentability{} problem is solved for the 1RDM \cite{parr_density-functional_1989}, the pure state \nrepresentability{} of the 1RDM poses already considerable challenges (see e.g.~\cite{klyachko_quantum_2006, schilling_pinning_2013}). For higher-order RDMs the problem is still widely open. For the 2RDM a method has been formulated to construct necessary \nrepresentability{} conditions given a discrete single-particle basis \cite{mazziotti_structure_2012}.\\
It has been shown that imposing \nrepresentability{} conditions in variational calculations of the ground state energy using the 2RDM substantially improves the obtained energies \cite{zhao_reduced_2004}. In the context of the so-called contracted Schr\"odinger equation \cite{valdemoro_contracted_1997} for approximate solutions of the ground state problem, the concept of purification of correlated RDMs has been introduced \cite{mazziotti_purification_2002,valdemoro_n_2000,alcoba_spin_2005}: After each iterative step within the contracted Schr\"odinger equation, the 2RDM is modified to preserve a certain set of \nrepresentability{} conditions and spin symmetries while maintaining its contraction consistency with the 1RDM. This is achieved by using the unitary decomposition of hermitian matrices \cite{auchin_characteristic_1983, sun_lie_1984, casida_geometry_1986}.\\
In time-dependent settings, attempts to close the BBGKY hierarchy on the second level and propagating the 2RDM self-consistently based on approximations of the three-paricle RDM (3RDM) faced several stability issues (see e.~g.~\cite{schafer-bung_correlated_2008,Akbari2012}). Stability could be restored for many parameter regimes of interest in terms of correlation energies and external driving within the time-dependent two-particle reduced density matrix (TD2RDM) method by applying purification to the 2RDM \cite{lackner_propagating_2015, lackner_high-harmonic_2017,donsa_nonequilibrium_2023,brezinova_dynamical_2024}. It has been shown that in these parameter regimes it is sufficient for stability to require the positive semidefiniteness of the 2RDM and the related two-hole RDM, which correspond to two \nrepresentability{} conditions often referred to as the D- and the Q-conditions, respectively. In \cite{lackner_high-harmonic_2017}, an efficient purification scheme was proposed that incorporates the D- and the Q-conditions of the 2RDM while maintaining contraction consistency with the 1RDM. However, this algorithm did not guarantee that other symmetries of the Hamiltonian associated with conserved quantities are preserved after the purification. Therefore, the algorithm was subsequently corrected to incorporate energy conservation \cite{joost_dynamically_2022} and the $\eta$-symmetry of the Fermi-Hubbard model \cite{brezinova_dynamical_2024}. In both cases, the preservation of these symmetries was achieved, but at the cost of modifying the 2RDM to such an extent that the iterative purification failed to converge in some cases, especially in the presence of large correlation energies \cite{donsa_nonequilibrium_2023}.\\
More recently, it has been observed that similar stability issues 
also affect non-equilibrium Green's function methods that scale linearly in time \cite{schlunzen_achieving_2020,joost_g1-g2_2020,joost_dynamically_2022, tuovinen_time-linear_2023, balzer_accelerating_2023, bonitz_accelerating_2024}. These methods involve equal-time-limits of Green's functions, which correspond to RDMs. Therefore, it is expected that the stability issues observed in closures of the BBGKY hierarchy are carried over.\\
In this paper, we propose a new purification algorithm applicable to general \prdm{p}s, which is able to incorporate several \nrepresentability{} conditions, while preserving the contraction to the lower \prdm{(p-1)} space and any number of conserved quantities. This is achieved by exploiting the geometric picture within the so-called unitary decomposition of hermitian matrices \cite{auchin_characteristic_1983, sun_lie_1984,mazziotti_advances_2007} and by applying the Hilbert-Schmidt inner product for projections within the convex space of RDMs. For the specific case of purification within the TD2RDM method and application to the quench dynamics of the Fermi-Hubbard model as in \cite{donsa_nonequilibrium_2023}, we show that this projective purification scheme is superior to those previously applied in terms of the number of iterative steps required for purification, the flexibility in incorporating conserved quantities, and the stability achieved. Parameter regimes of interaction strengths and quenches become accessible that were previously inaccessible due to convergence issues. Our method is relevant for purifications of RDMs and (equal-time limits of) Green's functions in both time-independent and time-dependent settings. It enhances the precision of these methods, as well as expands their applicability to previously unattainable parameter regimes.\\
The paper is structured as follows: We introduce our projective purification scheme in Sec.~\ref{sec:purification}. Based on a specific test case of a quenched Fermi-Hubbard model explained in Sec.~\ref{sec:testcase}, we demonstrate the performance of the projective purification in Sec.~\ref{sec:results}. We conclude in Sec.~\ref{sec:conclusions}. The appendices contain an outline of a generalization of our projective purification algorithm and more detailed derivations related to the specific test case examined.
\section{Reduced density matrices and the projective purification scheme}\label{sec:purification}
\newcommand{\complex}{\mathbb{C}}
\newcommand{\naturals}{\mathbb{N}}
\newcommand{\reals}{\mathbb{R}}
\newcommand{\ket}[1]{\lvert #1\rangle}
\newcommand{\bra}[1]{\langle#1\rvert}
\newcommand{\ketbra}[2]{\lvert #1\rangle\langle#2\rvert}
\newcommand{\pmonepartobj}{D_{1\ldots (p-1)}}
\newcommand{\ppartobj}{D_{1\ldots p}}
\newcommand{\pholeobj}{Q_{1\ldots p}}
\newcommand{\Npobj}{D_{1\ldots N}}

The goal of our projective purification scheme is to enforce \nrepresentability{} conditions onto reduced density matrices, while preserving observables related to symmetries of the Hamiltonian. The main structure of this algorithm is based on the earlier work of \cite{lackner_high-harmonic_2017, joost_dynamically_2022, brezinova_dynamical_2024}. However, significant changes have been made to how additional constraints related to symmetries of the Hamiltonian are implemented. We limit our discussion in the following to time-dependent problems. The algorithm is equally applicable to time-independent problems, when purifications of RDMs are necessary.\\
We consider a system of $N$ identical fermionic particles with the wavefunction $\ket{\Psi}$ and we assume that the many-body Hilbert space is spanned by Slater determinants formed from a finite number of one-particle orbitals $\phi_1,\ldots,\phi_r$, $r\in \naturals$. The \prdm{p} $\ppartobj\in\complex^{r^p\times r^p}$ is obtained from the expectation value of the following normal-ordered string of creation and annihilation operators:
\begin{equation}\label{eq:prdm_def}
(\ppartobj)^{i_1\ldots i_p}_{j_1\ldots j_p} = 
\bra{\Psi}
a_{j_1}^\dagger\ldots a_{j_p}^\dagger
a_{i_p}\ldots a_{i_1}
\ket{\Psi}.
\end{equation}
In complete analogy we define the $p$-hole reduced density matrix (\phrdm{p}) $\pholeobj{}$ as follows:
\begin{equation}\label{eq:phrdm_def}
(\pholeobj)^{i_1\ldots i_p}_{j_1\ldots j_p} = 
\bra{\Psi}
a_{i_1}\ldots a_{i_p}
a_{j_p}^\dagger\ldots a_{j_1}^\dagger
\ket{\Psi}.
\end{equation}
The operators are thereby all evaluated at the same time $t$. The \prdm{p} $\ppartobj{}(t)$ as well as the \phrdm{p} $\pholeobj{}(t)$ may thus be thought of as equal-time limits of the corresponding $p$-body Green's functions \cite{joost_dynamically_2022}.\\
Note that $\ppartobj{}$ and $\pholeobj{}$ are by no means independent. Instead, $\pholeobj{}$ may be computed in terms of $\ppartobj{}$ via the relation \cite{ruskai_n_1970, mazziotti_cse_inbook2007}
\newcommand{\antisym}{{\hat{\mathcal{A}}}}
\newcommand{\identity}{I}
\begin{align}\label{eq:formula_for_q}
\pholeobj{} = \sum_{q=0}^p (-1)^q \antisym D_{1\ldots q}\identity_{(q+1)\ldots p},
\end{align}
where $\antisym$ is the anti-symmetrization operator counting equivalent permutations only once, and $\identity_{(q+1)\ldots p}$ are identity matrices in the corresponding reduced spaces. The \nrepresentability{} conditions we will focus on in the following are the positive semidefiniteness of both $\ppartobj{}$ and $\pholeobj{}$, i.~e.~the $D$- and the $Q$-condition:
\begin{equation}\label{eq:d_and_q_condition}
\ppartobj{}\geq 0, \hspace{0.5cm} \pholeobj{}\geq 0.
\end{equation}
These two conditions enforced upon the \prdm{2} proved essential in stabilizing the TD2RDM method \cite{lackner_propagating_2015, lackner_high-harmonic_2017, donsa_nonequilibrium_2023, brezinova_dynamical_2024}. Generalizations to further \nrepresentability{} conditions are explained in App.~\ref{ap:generalization}. \\
A further class of constraints onto the matrix $\ppartobj{}$ arises in systems with fixed particle number. In that case, the \prdm{p} and the \prdm{(p-1)} are not independent. Instead, the \prdm{(p-1)} may be computed from $\ppartobj{}$ by taking the partial trace over the $p^{\text{th}}$ index:
\begin{equation}\label{eq:contraction_consistency}
\pmonepartobj{} = \frac{1}{N-p}\Tr_p \ppartobj{}.
\end{equation}
Typically, propagation errors appear first in higher-order objects, and it is natural to try to purify the \prdm{p} without altering the \prdm{(p-1)}. The proposed algorithm includes a way to do so. We call this additional condition the \specialterm{contraction consistency} (CC).\\
Finally, during the time evolution within the approximate method the matrix $\ppartobj{}$ has to satisfy further constraints that are not directly linked to its \nrepresentability{} but are related to conserved observables, i.~e.~to all operators $X_i$ that commute with the Hamiltonian. Even if the equations of motion for the \prdm{p} are sufficiently well approximated such that these observables are conserved, the purification of the \prdm{p} applied after propagation steps might destroy these symmetries if special care is not taken. The fixed value of the operators $X_i$ is determined by the initial state:
\begin{equation}\label{eq:constant_of_motion}
\Tr(\ppartobj{}X_i) = X_i(t=0).
\end{equation}
\begin{figure}
    \centering
    \includegraphics{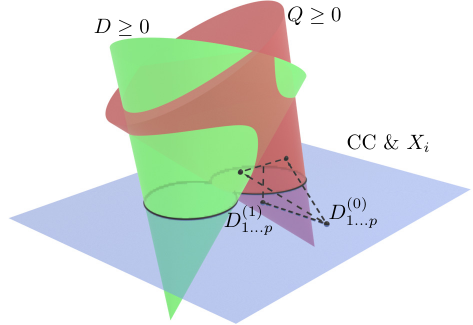}
    \caption{A sketch of the working principle of the projective purification. The defective \prdm{p} is projected onto the two cones corresponding to Eq.~\eqref{eq:d_and_q_condition}. The two results are then combined according to Eq.~\eqref{eq:defective_M_DQ} and projected onto the set of affine constraints resulting in Eq.~\eqref{eq:total_iteration_update}. This gives a new \prdm{p} $D_{1\ldots p}^{(1)}$, which is subjected to the same procedure resulting eventually in a purified $D_{1\ldots p}^{(k)}$ after a finite number of iteration steps $k$.}
    \label{fig:geometrical_sketch}
\end{figure}
Given a defective \prdm{p}, the goal of the purification algorithm is to restore the $D$- and the $Q$-conditions [Eq.~\eqref{eq:d_and_q_condition}] while retaining contraction-consistency [Eq.~\eqref{eq:contraction_consistency}] and the conservation of constants of motion [Eq.~\eqref{eq:constant_of_motion}]. This task can be visualized in a geometrical picture. In fact, both the $D$- and the $Q$-condition [Eq.~\eqref{eq:d_and_q_condition}] state that $\ppartobj{}$ lies in a corresponding affine pointed convex cone. The CC condition as well as any constants of motion require the $\ppartobj{}$ to lie in certain affine subspaces. The task achieved by the proposed algorithm is then to project a given \prdm{p} onto the intersection of all these sets (see Fig.~\ref{fig:geometrical_sketch}). This can be done with the help of the classical alternating projections algorithm \cite{bauschke_alternating_projections_1993}, consisting of alternating projections onto the individual convex sets. Instead of implementing the affine constraints directly, it is easier to restrict any change of $\ppartobj{}$ to the corresponding linear spaces. As long as the defective \prdm{p} $\ppartobj{}$, which is input to the algorithm, fulfills Eqs.~\eqref{eq:contraction_consistency}-\eqref{eq:constant_of_motion}, so will then the purified \prdm{p}. This requires that the approximate equations of motion are conserving, i.~e.~conserve all the required symmetries. The linear spaces corresponding to Eqs.~\eqref{eq:contraction_consistency}-\eqref{eq:constant_of_motion} are defined respectively via
\begin{equation}\label{eq:contraction_consistency_linear}
    \Tr_p\delta\ppartobj{} = 0,
\end{equation}
and
\begin{equation}\label{eq:conserved_quantities_linear}
\Tr\left(\delta\ppartobj{}X_i\right) = 0,
\end{equation}
where $\delta \ppartobj{}$ is any change that the purification algorithm enforces upon the \prdm{p} $\ppartobj{}$.\\
In practice, the projection can be achieved as follows: Given the \prdm{p} $D_{1\ldots p}^{(0)}$, we first compute the \phrdm{p} $Q_{1\ldots p}^{(0)}$ using Eq.~\eqref{eq:formula_for_q}. We then summarize both matrices within a vector
\begin{equation}
\mathcal{M}^{(0)} = \left(\begin{matrix}D_{1\ldots p}^{(0)}\\Q_{1\ldots p}^{(0)}\end{matrix}\right).
\end{equation}
This vector denotes now the initial value for the iteration described in the following. We first compute the defective component $D_{1\ldots p;\textnormal{def}}^{(k)}$ and $Q_{1\ldots p;\textnormal{def}}^{(k)}$ of each of the entries of $\mathcal{M}^{(k)}$, corresponding to that part of $\mathcal{M}^{(k)}$ violating the $D$- or the $Q$-condition.  At each step $k$ of the iteration, the defective component of each density matrix is computed by employing its spectral decomposition and keeping only negative eigenvalues:
\begin{equation}
D^{(k)}_{1\ldots p} = \underbrace{\sum_{g_i<0}g_i \ketbra{g_i}{g_i}}_{D^{(k)}_{1\ldots p;\textnormal{def}}} + \sum_{g_i\geq0}g_i \ketbra{g_i}{g_i}.
\end{equation}
We thus get
\begin{equation}\label{eq:defective_M}
\mathcal{M}^{(k)}_{\textnormal{def}} := \left(\begin{matrix}
D_{1\ldots p;\textnormal{def}}^{(k)}\\Q_{1\ldots p;\textnormal{def}}^{(k)}
\end{matrix}\right).
\end{equation}
Now simply subtracting the defective $\mathcal{M}_{\textnormal{def}}^{(k)}$ from $\mathcal{M}^{(k)}$ would break the relation Eq.~\eqref{eq:formula_for_q} between the \prdm{p} and the \phrdm{p}. Therefore, in the next step we must project the iteration update $\mathcal{M}_{\textnormal{def}}^{(k)}$ onto the linear space that corresponds to Eq.~\eqref{eq:formula_for_q}, in addition to the spaces corresponding to contraction consistency and conserved quantities [Eqs.~\eqref{eq:contraction_consistency_linear}-\eqref{eq:conserved_quantities_linear}]. Bearing in mind that we simultaneously enforce the contraction consistency, this linear constraint simply reads
\begin{equation}\label{eq:formula_for_dq}
 \delta Q_{1\ldots p} = (-1)^p \delta D_{1\ldots p}.
\end{equation}
We will now explain how the projections onto the constraints Eqs.~\eqref{eq:formula_for_dq},\eqref{eq:contraction_consistency_linear} and \eqref{eq:conserved_quantities_linear} can be attained. The first projection results from basic linear algebra and corresponds to simply setting
\begin{align}\label{eq:defective_M_DQ}
    \mathcal{M}_{\textnormal{def};DQ}^{(k)} &:= \frac{1}{2}\left(\begin{matrix}1&(-1)^p\\(-1)^p&1\end{matrix}\right)\mathcal{M}^{(k)}_{\textnormal{def}} \\\nonumber &=
    \frac{1}{2}\left(\begin{matrix}
        D_{1\ldots p;\textnormal{def}}^{(k)}+(-1)^pQ_{1\ldots p;\textnormal{def}}^{(k)}\\
        Q_{1\ldots p;\textnormal{def}}^{(k)}+(-1)^pD_{1\ldots p;\textnormal{def}}^{(k)}
    \end{matrix}\right).
\end{align}
\newcommand{\sumofdandq}{M}
In order to simplify further notation we denote the sum appearing in Eq.~\eqref{eq:defective_M_DQ} in the following by $\sumofdandq^{(k)}_{1\ldots p;\textnormal{def}}$
\begin{equation}
    \sumofdandq^{(k)}_{1\ldots p;\textnormal{def}} := D_{1\ldots p;\textnormal{def}}^{(k)}+(-1)^pQ_{1\ldots p;\textnormal{def}}^{(k)}.
\end{equation}
The projection onto the space defined by Eq.~\eqref{eq:contraction_consistency_linear} is effectively the projection onto the kernel of the partial trace map. As has been argued earlier (see e.g.~\cite{mazziotti_purification_2002,lackner_high-harmonic_2017}), this can be computed by employing the unitary decomposition \cite{sun_lie_1984,auchin_characteristic_1983, mazziotti_advances_2007,lackner_high-harmonic_2017}, giving a representation of any density matrix in terms of the sum of a contraction-free component (i.~e.~the kernel component) and another component which is orthogonal to the kernel with respect to the Hilbert-Schmidt inner product
\begin{equation}
\sumofdandq_{1\ldots p;\textnormal{def}} = \sumofdandq_{1\ldots p;\textnormal{def};K}+\sumofdandq_{1\ldots p;\textnormal{def};\perp}.
\end{equation}
This decomposition can be applied to the sum $\sumofdandq^{(k)}_{1\ldots p;\textnormal{def}}$ resulting in
\begin{equation}\label{eq:projection_onto_kernel}
\mathcal{M}_{\textnormal{def};DQ;K}^{(k)} := \frac{1}{2}\left(\begin{matrix}
        \sumofdandq^{(k)}_{1\ldots p;\textnormal{def};K}\\
        (-1)^p\sumofdandq^{(k)}_{1\ldots p;\textnormal{def};K}
    \end{matrix}\right).
\end{equation} 
Finally, we want to explain how to implement the projection onto the spaces defined by Eq.~\eqref{eq:conserved_quantities_linear}. In earlier work \cite{joost_dynamically_2022,donsa_nonequilibrium_2023,brezinova_dynamical_2024}, this was done by analyzing which matrix elements of $D_{1\ldots p}$ contribute to the computation of the respective observables. All these matrix elements would then have been set to zero in order to achieve this projection. However, this corresponds to a projection onto a much smaller space than the one described by Eq.~\eqref{eq:conserved_quantities_linear}. As such it imposes a much more stringent constraint onto to the purification, leading to slower convergence of the algorithm. In some cases, the additional constraints may even conflict with any of the other conditions Eqs.~\eqref{eq:d_and_q_condition}-\eqref{eq:contraction_consistency} resulting in a failure to converge at all.\\
In our new algorithm the projection is instead achieved as follows: We first project all the operators $X_1,\ldots X_n$ onto the kernel of the partial trace by using again the unitary decomposition. This ensures that the projection defined in the following does not conflict with the earlier projection Eq.~\eqref{eq:projection_onto_kernel}. We then orthonormalize the set of all operators $\{X_{1;K},\ldots,X_{n;K}\}$ with respect to each other using the Gram-Schmidt-routine and the Hilbert-Schmidt inner product of matrices. This gives rise to a new set of operators $\{Y_1,\ldots Y_n\}$. With these operators at hand, we may implement the third projection as follows:
\begin{equation}
\label{eq:yconserving_projection}
\sumofdandq_{1\ldots p;\textnormal{def};K;\mathcal{X}} := \sumofdandq_{1\ldots p;\textnormal{def};K} - \sum_{i=1}^n\Tr(\sumofdandq_{1\ldots p;\textnormal{def};K}Y_i)Y_i.
\end{equation}
This is the orthogonal projection with respect to the Hilbert-Schmidt inner product onto the admissible space of iteration updates that have zero expectation in the space spanned by $\{X_1,\ldots,X_n\}$. As such this projection enforces the conservation of these quantities with the least possible change (measured in terms of the Hilbert-Schmidt metric) to the iteration update. In fact, evaluating this projection amounts to the removal of only one degree of freedom per operator $X_i$.
Altogether, the iteration update becomes
\begin{equation}\label{eq:total_iteration_update}
\mathcal{M}^{(k)}_{\textnormal{def};DQ;K;\mathcal{X}} := 
\frac{1}{2}\left(
\begin{matrix}
\sumofdandq_{1\ldots p;\textnormal{def};K;\mathcal{X}}\\
(-1)^p\sumofdandq_{1\ldots p;\textnormal{def};K;\mathcal{X}}
\end{matrix}\right).
\end{equation}
And one iteration of the algorithm reads
\begin{equation}
\label{eq:M_iter}
\mathcal{M}^{(k+1)} = \mathcal{M}^{(k)} - \alpha\mathcal{M}^{(k)}_{\textnormal{def};DQ;K;\mathcal{X}}.
\end{equation}
Here, we introduced an additional steering parameter $\alpha>0$, which may be used to control the algorithm's convergence.
\section{Test case}\label{sec:testcase}
We now study the performance of the projective purification algorithm within the TD2RDM method. Our criteria for assessing the quality of the projective purification scheme are the ability to stabilize the propagation, the convergence of the computed observables with the time steps at which the purification is performed, and the convergence of the iteration within the purification algorithm itself. As a platform for benchmarking, we use the quench dynamics in the paradigmatic Fermi-Hubbard model, which was studied within the TD2RDM method in \cite{donsa_nonequilibrium_2023} using previously applied purification schemes \cite{lackner_high-harmonic_2017, joost_dynamically_2022}. In the following, we will assume a unit system in which $e$, $\hbar$, and $m_e$ are all equal to 1.\\
Specifically, we consider a system described by the following one-dimensional Fermi-Hubbard Hamiltonian including an external potential given by
\begin{equation}\label{eq:hubbard_hamiltonian}
    H = - J\sum_{\langle i,j \rangle,\sigma} a_{i \sigma} ^{\dagger} a_{j \sigma} + U\sum_i n_{i\uparrow}n_{i\downarrow}+\sum_{i,\sigma}V_i(t)n_{i\sigma},
\end{equation}
where $J$ is the hopping amplitude, $\langle i,j \rangle$ are the indices of each possible nearest-neighbor pair, $U$ is the interaction strength, and  $V_i(t)$ is a time-dependent potential. We assume hard wall boundary conditions at the border of the system. The number of sites is denoted as $M_s$, the number of particles $N$ is set to half-filling, i.~e.~$N=M_s$, and the number of spin-up electrons is equal to the number of spin-down electrons. In all that follows we will set 
\begin{equation}\label{eq:quench_potential}
    V_i(t)=\theta(-t)\frac{V^2}{2}\left(i-\frac{M_s+1}{2}\right)^2,
\end{equation}
which amounts to simulating the free dynamics of an initially harmonically trapped system. The initial state is taken as the ground state of the system with harmonic trap. \\
Within the \tdtwordm{} method, the dynamics of the \prdm{2} is governed by the second element of the BBGKY-hierarchy \cite{bonitz_accelerating_2024,lackner_propagating_2015,lackner_high-harmonic_2017}. The corresponding equation reads
\begin{align}\label{eq:eom}
i\pdv{t}D_{12} =& [h_1 + h_2 + W_{12},D_{12}] \\\nonumber 
 &+ \Tr_{3}[W_{13}+W_{23},D_{123}].
\end{align}
Here, $h_i$ comprises all single-particle operators entering the Hamiltonian Eq.~\eqref{eq:hubbard_hamiltonian}, while $W_{ij}$ denotes the interaction. The matrix elements of $h_1$ and $W_{12}$ are given in the present case by
\begin{align} \label{eq:matrix_elements}
    \langle i\sigma'|h_1|j\sigma\rangle &= -J\delta_{j}^{i-1}\delta_{\sigma}^{\sigma'} - J\delta_{j}^{i+1}\delta_{\sigma}^{\sigma'} \nonumber \\
    \langle i_1\sigma_1'i_2\sigma_2' |W_{12} |j_1\sigma_1j_2\sigma_2\rangle &= U\delta_{j_1}^{i_1}\delta_{j_2}^{i_2}\delta_{j_1j_2}\delta_{\sigma_1}^{\sigma_1'}\delta_{\sigma_2}^{\sigma_2'}(1-\delta_{\sigma_1\sigma_2}).
\end{align}
In order to close the system of equations Eq.~\eqref{eq:eom} and make it explicit in $D_{12}$, it is necessary to approximate $D_{123}$ in terms of $D_{12}$. This is done by employing approximate reconstruction functionals based on the cumulant decomposition of reduced density-matrices \cite{mazziotti_approximate_1998}. In this paper, we made use of the Valdemoro reconstruction functional \cite{colmenero_approximating_1993,Valdemoro2007} (for a review see e.g.~ \cite{mazziotti_advances_2007}) enforcing contraction consistency of the \prdm{3} \cite{lackner_propagating_2015,lackner_high-harmonic_2017}. For more details about the reconstruction within the present test case we refer to \cite{brezinova_dynamical_2024,donsa_nonequilibrium_2023}.\\
Eq.~\eqref{eq:eom} becomes approximate as soon as one replaces $D_{123}$ with the reconstruction such that the dynamics of 
$D_{12}$ will in general deviate from the exact dynamics. This can lead to the \prdm{2} losing its \nrepresentability{}, which, if not taken care of, gives rise to the problem of instabilities \cite{Akbari2012}. In order to solve this problem, the projective purification algorithm as described in Sec.~\ref{sec:purification} is applied between any two propagation steps.\\
In order to be able to perform large parameter scans and still compare to exact results, we choose a rather small system size of $M_s=6$. The ground state of this system fulfills $S^2=0$ \cite{Lieb1989}. This fact may be exploited to decompose the \prdm{2} $D_{12}$ into a spin-singlet component $D_{12}^S$ and a spin-triplet component $D_{12}^T$ and to apply the purification scheme separately to both components \cite{alcoba_spin_2005, lackner_propagating_2015}. The two components are given by
\begin{equation}
    (D_{12}^S)^{i_1i_2}_{j_1j_2} = \bra{\psi}
      a_{j_1j_2;S}^\dag a_{i_1i_2;S}
    \ket{\psi}
\end{equation}
and
\begin{equation}
    (D_{12}^T)^{i_1i_2}_{j_1j_2} = \bra{\psi}
      a_{j_1j_2;T}^\dag a_{i_1i_2;T}
    \ket{\psi},
\end{equation}
where 
\begin{equation}
    a_{j_1j_2;T/S}^\dag = \frac{1}{\sqrt{2}} (a_{j_1\uparrow}^{\dagger}a_{j_2\downarrow}^{\dagger}\pm a_{j_1\downarrow}^{\dagger}a_{j_2\uparrow}^{\dagger}).
\end{equation}
Just like in the general description of the algorithm, we only enforce the $D$- and the $Q$-condition [Eq.~\eqref{eq:d_and_q_condition}] since these two have proven to stabilize the time propagation of the \tdtwordm{} method \cite{lackner_propagating_2015, lackner_high-harmonic_2017, donsa_nonequilibrium_2023}. We have observed empirically that the so-called $G$-condition \cite{garrod_reduction_1964,mazziotti_advances_2007}, i.~e., the positive semidefinitness of the particle-hole RDM, is automatically satisfied when the $D$- and the $Q$-conditions are satisfied, such that it does not have to be enforced in addition. At this point we would like to mention that the computationally expensive full diagonalization of the \prdm{2} and \phrdm{2} is not necessary within the purification, since the space spanned by negative eigenvalues can be computed with more efficient methods, such as the Arnoldi iteration algorithm \cite{arnoldi_principle_1951}. Generalizations to further \nrepresentability{} conditions are also possible and the path towards them is demonstrated in App.~\ref{ap:generalization}. However, from a computational point of view, it is numerically increasingly expensive to implement further \nrepresentability{} constraints, such that within a time propagation one is essentially limited to a few of them.\\
As the Hamiltonian Eq.~\eqref{eq:hubbard_hamiltonian} conserves the number of particles, we also ensure contraction consistency [Eq.~\eqref{eq:contraction_consistency}].
The remaining constraints come from physically conserved observables. Since the system at hand is closed, energy is a conserved quantity. The contraction consistency already ensures the conservation of single-particle energy contributions, which is why as an additional conserved operator we only have to include the interaction $W$. The expectation value of the interaction energy may be computed from $D_{12}$ using
\begin{equation}\label{eq:energy_index_formula}
   \langle W\rangle=\frac{U}{2}\sum_i \left[ D_{12}^S\right] ^{ii}_{ii}.
\end{equation}
$D_{12}^T$ does not contribute to the interaction energy. The first operator $X_1$ from Eq.~\eqref{eq:constant_of_motion} appears therefore only in the purification of $D_{12}^S$ and is explicitly given by
\begin{equation}
    (X_1)^{i_1i_2}_{j_1j_2} = \frac{U}{2}\delta^{i_1}_{j_1}\delta^{i_1}_{j_2}\delta^{i_2}_{j_1}.
\end{equation}
For the purification, only the normalized contraction-free component $Y_1$ of $X_1$ is used
\begin{equation}
\label{eq:index_formula_for_y1}
(Y_1)^{i_1i_2}_{j_1j_2} = \frac{(M_s+1)\delta^{i_1}_{j_1}\delta^{i_1}_{j_2}\delta^{i_2}_{j_1}-\delta^{i_1}_{j_1}\delta^{i_2}_{j_2}-\delta^{i_1}_{j_2}\delta^{i_2}_{j_1}}{\sqrt{(M_s-1)M_s(M_s+1)}}.
\end{equation}
The Fermi-Hubbard model also exhibits the $\eta$-symmetry \cite{yang__1989}. This symmetry is associated with the conservation of the expectation value $\langle \eta^+ \eta^- \rangle$, where the operators $\eta^+$ and $\eta^-$ are defined via:
\begin{equation}
 \eta^+=\sum_j (-1)^j a^\dagger_{j \downarrow} a^\dagger_{j \uparrow}, \hspace{0.5cm}
 \eta^-=(\eta^+)^\dag.
\end{equation}
In terms of $D_{12}$, the expectation value $\langle \eta^+ \eta^- \rangle$ is evaluated as follows:
\begin{equation}\label{eq:eta_index_formula}
    \langle\eta^+ \eta^- \rangle = \frac{1}{2}\sum_{i,j}(-1)^{i+j}\left[D_{12}^S\right] ^{jj}_{ii}.
\end{equation}
Again, $D_{12}^T$ does not contribute to the expectation value and the purification of the triplet component of $D_{12}$ conserves this symmetry trivially. The second operator $X_2$ from Eq.~\eqref{eq:constant_of_motion} becomes thus
\begin{equation}
    (X_2)^{i_1i_2}_{j_1j_2} = \frac{(-1)^{i_1+j_1}}{2}\delta^{i_1i_2}\delta_{j_1j_2}.
\end{equation}
As is described in Sec.~\ref{sec:purification}, the kernel component of this operator must be orthogonalized to $Y_1$ to yield $Y_2$. The result of this is
\begin{equation}
    \label{eq:index_formula_for_y2}
    (Y_2)^{i_1i_2}_{j_1j_2} = \frac{(-1)^{i_1+j_1}\delta^{i_1i_2}\delta_{j_1j_2}-\delta^{i_1}_{j_1}\delta^{i_1}_{j_2}\delta^{i_2}_{j_1}}{\sqrt{(M_s-1)M_s}}.
\end{equation}
The conservation of $S^2=0$ is incorporated partly within the parametrization of the density matrix and partly by the requirement of contraction-consistency with respect to the \prdm{1}. Therefore, it does not appear as a separate conserved operator $X_3$. Hence, the two operators $Y_1$ and $Y_2$ may now be plugged into Eq.~\eqref{eq:yconserving_projection} to yield the final expressions for $M^{(k)}_{\textnormal{def};DQ;K;\mathcal{X}}$ needed in Eq.~\eqref{eq:M_iter}. Their specific form can be found in App.~\ref{ap:iteration}.
\section{Results}\label{sec:results} 
\begin{figure}
\includegraphics[width=\columnwidth]{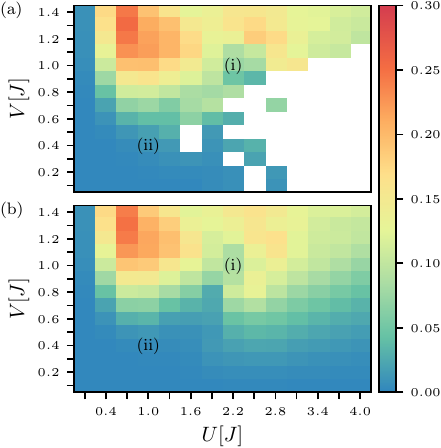}
\caption{(a) The error in the density fluctuations as defined in Eq.~\eqref{eq:n1_error} for different parameters $U$ and $V$ in a computation employing the previously used purification algorithm \cite{brezinova_dynamical_2024}. White pixels correspond to computations that were not converged with respect to the threshold given in Eq.~\eqref{eq:convergence}. The labels (i) and (ii) refer to the two exemplary systems investigated in more detail in Figs.~\ref{fig:defectOverTime}-\ref{fig:purificationConvergence}.
(b) The same data and labels as in (a), here in a computation using the new projective purification. No problems with convergence appeared in this simulation.} 
\label{fig:dn1_error_scan}
\end{figure}
\begin{figure}
\includegraphics[width=\columnwidth]{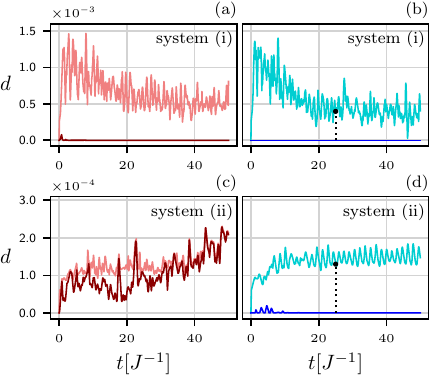}
\caption{The defect Eq.~\eqref{eq:def_defect} as a function of time for the two systems marked in Fig.~\ref{fig:dn1_error_scan} as (i) [subplots (a) and (b)] and (ii) [subplots (c) and (d)] for both the previous purification scheme [subplots (a) and (c)] and the projective purification [subplots (b) and (d)]. Every subplot contains two curves, the lighter one representing the value of the defect before applying the purification, and the darker one the respective value after at most 100 iterations of the purification algorithm. In subplots (b) and (d) the input-\prdm{2} used for the convergence analysis in Fig.~\ref{fig:purificationConvergence} is marked with a black dot on top of a dashed line.}
\label{fig:defectOverTime}
\end{figure}
\begin{figure} 
\includegraphics[width=\columnwidth]{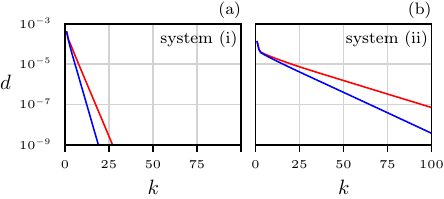}
\caption{The defect Eq.~\eqref{eq:def_defect} at one time-step as a function of the iteration index $k$ within the purification scheme. The subplots (a) and (b) correspond to the systems (i) and (ii) marked in Fig.~\ref{fig:dn1_error_scan}. The red curve shows the convergence of the purification when using the old implementation, while the blue curve corresponds to the projective purification. The input \prdm{2} for both purification algorithms was computed by propagating the initial states up to $t=25 J^{-1}$ using the projective purification. The defect of the input \prdm{2} is marked in Fig.~\ref{fig:defectOverTime} (b) and (d) with a black dot on top of a dashed line.}
\label{fig:purificationConvergence}
\end{figure}

In this section, we compare the performance of the newly proposed projective purification algorithm with the implementation used in previous works \cite{joost_dynamically_2022, donsa_nonequilibrium_2023, brezinova_dynamical_2024}. For this purpose, the purification algorithm described in Sec.~\ref{sec:purification} is applied to the \tdtwordm{} simulation of the Fermi-Hubbard model described in Sec.~\ref{sec:testcase}. Further details on the numerical parameters used to produce the presented data are discussed in App.~\ref{ap:num_details}. The previous and present purification algorithms differ from each other by the way the conservation of constants of motion is achieved. Instead of projecting out the components of the iteration update $D_{12}$ in the direction of the operators $Y_i$ [Eq.~\eqref{eq:yconserving_projection}], the previous implementation would set any element of the iteration update appearing in Eqs.~\eqref{eq:energy_index_formula} and \eqref{eq:eta_index_formula} to zero. Doing so results in a much more stringent constraint on the purification algorithm and is therefore expected to yield slower convergence. In some cases, the previous implementation might even fail to purify a given density matrix. If the purification routine fails or converges too slowly, the outcome of the simulation depends strongly on the frequency of how often the purification is applied. Since the purification is applied between any two propagation steps whenever the defective part of $\mathcal{M}_{12}$ is non-vanishing, this frequency is directly linked to the numerical time step $dt$. Hence, a failure of the purification to converge can lead to results that are not converged with respect to the numerical time step $dt$. This is exactly where the projective purification leads to a striking advantage compared to the previous algorithm. Using the same convergence criterion as in \cite{donsa_nonequilibrium_2023}, i.~e.~  
\begin{equation}
\label{eq:convergence}
    \frac{1}{T}\int^T_0 dt \left|n_1^{dt}(t)-n_1^{dt'}(t) \right| < 5\cdot 10^{-3},
\end{equation}
where $n_1(t)$ is the density at first site chosen as a representative observable, and computed for two different time steps $dt$ and $dt'=0.5dt$, we observe that a large range of previously inaccessible parameters $V$ and $U$ become accessible, see Fig.~\ref{fig:dn1_error_scan}. Note that results for substantially larger interaction strengths $U$ can now be calculated without convergence problems compared to the results obtained in \cite{donsa_nonequilibrium_2023}. These parameter ranges were inaccessible exactly because the purification did not converge even with a large amount of maximal iterations $k_\text{max}$ allowed. There are slight deviations of Fig.~\ref{fig:dn1_error_scan} (a) compared to the corresponding Fig.~12 in \cite{donsa_nonequilibrium_2023}, which result from the fact that only the conservation of energy has been taken into account within the purification in \cite{donsa_nonequilibrium_2023}. The conservation of $\eta^+\eta^-$ within the \tdtwordm{} simulations of the Fermi-Hubbard model where introduced later in \cite{brezinova_dynamical_2024}. Constraining the purification to conserve multiple observables renders the purification algorithm even more difficult to converge within the previous purification scheme and leads to an even larger inaccessible parameter range in Fig.~\ref{fig:dn1_error_scan} (a).
However, as is visible in Fig.~\ref{fig:dn1_error_scan} (b), with the projective purification introduced here we obtain converged results for all parameter pairs analyzed.\\
The quantity visualized in Fig.~\ref{fig:dn1_error_scan} is the averaged deviation of the occupation number of the left-most site from exact results
\begin{equation}
\label{eq:n1_error}
    \overline{\delta n_1}=\frac{\int_0^Tdt\left| n_1(t) - n_1^\text{exact}(t) \right|}{\int_0^Tdt \, n_1^\text{exact}(t)}.
\end{equation}
It can thus be seen from Fig.~\ref{fig:dn1_error_scan} that the TD2RDM is in good agreement with the exact results within the region hitherto inaccessible. Comparing the results within TD2RDM for the two different purification schemes, we observe that the results do not depend on the purification scheme as long as the purification procedure is sufficiently well behaved. Slight deviations between Fig.~\ref{fig:dn1_error_scan} (a) and (b) are only present in parameter regions that are close to the white pixels in Fig.~\ref{fig:dn1_error_scan} (a), i.~e.~parameter regions where the previously applied purification starts to struggle to produce convergent results.\\
We now want to analyze in detail, for a few specific examples marked in Fig.~\ref{fig:dn1_error_scan} (a) and (b), to what extent the projective purification is more efficient compared to the previous algorithm in restoring the required \nrepresentability{} conditions. As a figure of merit we use the number of iterations $k$ needed to restore \nrepresentability{} without touching all the other symmetries discussed in Sec.~\ref{sec:purification}. In cases where the purification does not converge after a maximum number of iterations set to $k_\text{max} = 100$, we take the remaining defect $d$
\begin{equation}
\label{eq:def_defect}
d:=-\min\left(\min_{g_i<0}g_i , \, \min_{q_i<0} q_i \right)
\end{equation}
as a figure of merit. In the above formula, $\{g_i\}_i$ and $\{q_i\}_i$ denote the sets of eigenvalues of $D_{12}$ and $Q_{12}$, respectively.\\
By looking at $d$ for different systems as a function of time (Fig.~\ref{fig:defectOverTime}), one can make several observations. For some parameters, both purification schemes manage to enforce the \nrepresentability{} of $D_{12}$ and $Q_{12}$ at each time step. An example of such a system is given by the parameters $U=2.2J$ and $V=1.0J$, system (i) in Fig.~\ref{fig:dn1_error_scan}. In these cases, both purification schemes may be applied without significant influence on the results, see Fig.~\ref{fig:defectOverTime} (a) and (b).
In other cases, the previous implementation fails to purify the \prdm{2} within the maximal amount of allowed iterations $k_\text{max}$. An example of such a system is given by $U=1.0J$ and $V=0.4J$ [system (ii)]. While the defect is still small, and importantly smaller than the convergence criterion Eq.~\eqref{eq:convergence}, it will accumulate over time. Especially, after around $t=35J^{-1}$, the purification scheme ceases to have any meaningful mitigation effect, leading to a steady increase in $d$, which could lead to a failure of convergence and numerical stability if propagated further. For the same system, the projective purification succeeds in reducing the defect to zero at almost all time steps and thereby stabilizes the value of the defect before another propagation and purification step is applied to an amount still manageable for the scheme. For the  parameter combinations which are not converged in Fig.~\ref{fig:dn1_error_scan}, the previously used purification is incapable of mitigating $d$ right from the start.\\
While the advantage of using the projective purification is most pronounced for systems with large $U$ and small to moderate $V$, even for those systems where both purification algorithms are applicable, the newly proposed projective purification offers an advantage over the previously used algorithm. In these cases, it manages to reduce the defect to zero after significantly fewer iterations. This is visualized in Fig.~\ref{fig:purificationConvergence}.
Here, the same two exemplary systems are analyzed and the reduction of the defect by subsequent iterations is shown. In order to have a direct comparison between the two algorithms, the same input \prdm{2} was used in both cases, taken from the \tdtwordm{} computation using the  projective purification up until the time $t=25 J^{-1}$. It is clearly visible that the convergence is faster for the projective purification for both systems, even though the reduction of the defect with the iteration step $k$ is exponential in both cases [except for the sharper decrease during the first few iteration steps for system (ii)]. Note that, even though being significantly less efficient than the projective purification, the previous purification reduces the defect by several orders of magnitude even for the system with small $V$ [system (ii)]. This seems contradictory to Fig.~\ref{fig:defectOverTime}, where for the weakly excited system [system (ii)] at $t=25 J^{-1}$, the previous algorithm leads to a decrease of the defect by only a few percent. This is a direct consequence of the accumulation of the error over time, since the input \prdm{2} for Fig.~\ref{fig:defectOverTime} (c) originates entirely from a computation using the previous purification, while in Fig.~\ref{fig:purificationConvergence} (b) the projective purification is used until $t=25 J^{-1}$ and only then is the previous purification scheme applied. The fact that the previous implementation is still able to locally purify \prdm{2}s from a computation using the new projective purification hints to the fact that the projective purification might contribute to restoring also other \nrepresentability{} constraints not caught by the defect measure $d$. 
\section{Conclusions}\label{sec:conclusions}
In conclusion, we have presented a projective purification scheme to restore \nrepresentability{} conditions in non-$N$-representable reduced density matrices (RDMs). The projective purification scheme presented here is applicable to arbitrary RDMs and a wide variety of \nrepresentability{} conditions, while maintaining the contraction consistency between successive orders of RDMs and the conservation of observables related to symmetries of the Hamiltonian. The purification scheme is firmly rooted in the geometric representation of \nrepresentability{} and the symmetries of the Hamiltonian. It employs the alternating projections algorithm with the Hilbert-Schmidt inner product for matrices. We have demonstrated the application of the projective purification to the specific case of the time-dependent two-particle reduced density matrix (TD2RDM) method, where purifications of the two-particle reduced density matrix (2RDM) are necessary to obtain stable solutions in several parameter regimes of interaction strengths and external driving. Using the quench dynamics of the one-dimensional Fermi-Hubbard model as a benchmark system, we have shown that the present projective purification scheme greatly expands the accessible parameter regimes in terms of interactions and quench strengths compared to previous purification schemes. The projective purification safely reaches the required \nrepresentability{} conditions in all parameter regimes investigated, even in regimes where the previous purification schemes fail. In regimes where both the previous and the present purification schemes are applicable, the number of iterations required to restore the relevant \nrepresentability{} conditions is significantly smaller than within previous purification schemes. This is attributed to the fact that the present algorithm minimizes the amount of modifications to the RDM while still restoring or preserving all relevant symmetries.\\
Our purification scheme is relevant to all time-dependent and time-independent methods, which use reduced objects that might lose \nrepresentability{} within self-consistent applications of the method. These methods include RDM methods and non-equilibrium Green's function methods with linear scaling in time. Our projective purification algorithm might also find applications in the realm of quantum error mitigations \cite{cai_quantum_2023} relevant for quantum chemistry computations on noisy intermediate scale quantum computers (see e.~g.~\cite{rubin_application_2018, mcardle_quantum_2020,obrien_purification-based_2023,avdic_fewer_2024}).
%
\section*{Acknowledgements}
This research was funded by the Austrian Science Fund (FWF) grant P 35539-N, as well as by the FWF 10.55776/COE1. Calculations were performed on the Vienna Scientific Cluster (VSC4). 
\appendix
\section{Generalization to other \nrepresentability{} constraints}\label{ap:generalization}
In Sec.~\ref{sec:purification} we presented the purification algorithm for the case when only two \nrepresentability{} constraints are imposed: The $D$- and the $Q$-condition. However, the generalization to other \nrepresentability{} constraints is straightforward from a mathematical point of view and will be the content of this appendix. From a numerical point of view, high-order \nrepresentability{} conditions may still be unfeasible to implement due to their large polynomial scaling with the number of orbitals in the single-particle basis.\\
The set of $N$-representable \prdm{p}s has been intensely studied by Mazziotti, who managed to characterize it in terms of a sequence of positivity constraints \cite{Mazziotti2012, mazziotti_significant_2012}. These constraints may be organized in form of a hierarchical structure, where constraints on higher \prdm{p}s imply those on lower \prdm{p}s by constructing appropriate convex combinations.\\
An important subset of these constraints are those that can be expressed as the condition of positive semidefiniteness of some matrices $M\in\complex^{k\times k}$, $k\in\naturals$ depending directly and in an affine way on the \prdm{p}:
\begin{equation}\label{eq:constraint_def}
    M = f[\ppartobj{}] = K + L[\ppartobj{}] \geq 0
\end{equation}
Here, $K\in\complex^{k\times k}$ is some fixed matrix and $L:\complex^{r^p\times r^p}\rightarrow \complex^{k\times k}$ is a linear map.\\
An example is clearly the positive semidefiniteness of $\ppartobj{}$ itself, where $K = 0$ and $L=\id$, as well as the positive semidefiniteness of the corresponding \phrdm{p} $\pholeobj{}$. There are, however, many more such constraints.
These constraints are exactly those that can readily be included in a generalized version of the purification scheme proposed in Sec.~\ref{sec:purification}. 
\\In the following, we will explain how to do so.
Every constraint $f_1,\ldots,f_n$, $n\in\naturals$ of the form Eq.~\eqref{eq:constraint_def} defines a pointed convex cone in the space of \prdm{p}s. The goal of the purification scheme is then the projection of $\ppartobj{}$ onto the intersection of all these cones. This is achieved through an iterative procedure: Given the \prdm{p} $\ppartobj{}$, we first compute all corresponding matrices $M_i = f_i[\ppartobj{}]$ and align them, together with $\ppartobj{}\equiv M_0$, in a large vector
\begin{equation}
\mathcal{M} := 
\left(\begin{matrix}
\ppartobj{}\\M_1\\\vdots\\M_n
\end{matrix}\right).
\end{equation}
Note that the matrices $M_0,\ldots,M_n$ need not be of the same size. The vector $\mathcal{M}$ is thus not a complex matrix consisting of separate blocks, but a vector in the Hilbert space $C^{k_0\times k_0}\oplus \ldots \oplus C^{k_n\times k_n}$, where $k_i$ is the dimension of $M_i$.\\
We may then project every entry of this vector independently onto its corresponding cone. Since this cone is simply the cone of positive semidefinite matrices, the projection can be done by computing the negative eigenpairs and subtracting the corresponding projectors from $M_i$. We denote the results by $M_{i,\textnormal{pos}}$ and the subtracted (defective) part by $M_{i,\textnormal{def}}$, $i=1,\ldots,n$. Altogether, this amounts to a projection of $\mathcal{M}$, the result of which we denote by $\mathcal{M}_\textnormal{pos}$. Accordingly, we will use $\mathcal{M}_{\textnormal{def}}$ to refer to the difference $\mathcal{M}-\mathcal{M}_\textnormal{pos}$. Simultaneously, we want to introduce the notation for the projection $\pi_{\textnormal{def}}(\mathcal{M}) := \mathcal{M}_{\textnormal{def}}$.\\
Now, clearly, the new matrices $M_{i,\textnormal{pos}}$ will in general not satisfy the defining equality in Eq.~\eqref{eq:constraint_def} anymore. This can be repaired as follows: Since $M_0=\ppartobj{}$, the constraints $M_i = f_i[\ppartobj{}]$ are nothing but a series of affine constraints onto $\mathcal{M}$:
\begin{equation}
  L_i[M_0] - M_i = -K_i.
\end{equation}
We can enforce these constraints by making sure that the difference $\mathcal{M}_{\textnormal{def}}=\mathcal{M}-\mathcal{M}_\textnormal{pos}$  satisfies the corresponding linear constraints:
\begin{equation}
  L_i[M_{0,\textnormal{def}}] - M_{i,\textnormal{def}} = 0,\,\,i=1,\ldots,n.
\end{equation}
All constraints can be summarized in the following matrix equation featuring an $n\times(n+1)$-matrix of linear maps:
\begin{equation}\label{eq:matrix_eq}
\setlength\arrayrulewidth{0.2pt}
\left(\begin{matrix}
  L_1    & -\id &      &        &      \\
  L_2    &      & -\id &        &      \\
  \vdots &      &      & \ddots &      \\
  L_n    &      &      &        & -\id \\
\end{matrix}\right)
\mathcal{M}_{\textnormal{def}} = 0.
\end{equation}
It is useful to summarize the linear maps $L_i$ in an $n$-dimensional vector as well:
\begin{equation}
\mathcal{L} = \left(\begin{matrix}
L_1\\L_2\\\vdots\\L_n
\end{matrix}\right),
\end{equation}
since then Eq.~\eqref{eq:matrix_eq} simply reads:
\begin{equation}
\left(\begin{matrix}\mathcal{L}&\,\,-\id_n \end{matrix}\right)
\mathcal{M}_{\textnormal{def}} = 0.
\end{equation}
Here, we have introduced the symbol $\id_{n}$ for a diagonal $n\times n$-matrix containing only identities.
We thus search for the projection of $\mathcal{M}_{\textnormal{def}}$ onto the kernel of $\left(\begin{matrix}\mathcal{L}&\,\,-\id_n \end{matrix}\right)$. Using well-established formulae for the Moore-Penrose pseudoinverse, this $(n+1)\times(n+1)$-dimensional projection matrix can be expressed as:
\begin{equation}
\label{eq:major_projection}
\mathcal{P} = \id_{n+1} - 
\left(\begin{matrix}
\mathcal{L}^T\\
-\id_{n}\end{matrix}\right)
\left(\id_{n} + \mathcal{L}\mathcal{L}^T\right)^{-1}
\left(\begin{matrix}
\mathcal{L}&\,\,-\id_{n} \end{matrix}\right).
\end{equation}
The inversion of the matrix $(\id_n+\mathcal{L}\mathcal{L}^T)$ might at first glance seem to pose a problem because of vanishing eigenvalues. However, it can be obtained by making use of the Woodbury matrix identity, resulting in:
\begin{equation}
\label{eq:woodbury}
    \left(\id_n+\mathcal{L}\mathcal{L}^T\right)^{-1} = \id_n - \mathcal{L}{\underbrace{(\id+\mathcal{L}^T\mathcal{L})}_{=:\,C}}^{-1}\mathcal{L}^T.
\end{equation}
In order to make the following formulas more readable, we introduce a new symbol for the linear map $C := \id + \mathcal{L}^T\mathcal{L}$.
Inserting Eq.~\eqref{eq:woodbury} into Eq.~\eqref{eq:major_projection} and distributing terms gives:
\begin{align}
    \mathcal{P} =& \id_{n+1}-
    \left(\begin{matrix}
        \mathcal{L}^T\mathcal{L} & -\mathcal{L}^T \\
        -\mathcal{L}             &          \id_n  
    \end{matrix}\right) \\\nonumber
    &+\left(\begin{matrix}
        \mathcal{L}^T\mathcal{L}\\
                    -\mathcal{L}
    \end{matrix}\right) 
    C^{-1} \left(\begin{matrix}\mathcal{L}^T\mathcal{L}&-\mathcal{L}^T\end{matrix}\right).
\end{align}
Using the fact that 
\begin{equation}
    \mathcal{L}^T\mathcal{L}\,C^{-1} =
    C^{-1}\,\mathcal{L}^T\mathcal{L} = \id-\,C^{-1},
\end{equation}
we arrive at:
\begin{align}
    \mathcal{P} =&\id_{n+1}-
    \left(\begin{matrix}
        \mathcal{L}^T\mathcal{L} & -\mathcal{L}^T \\
        -\mathcal{L}             &          \id_n  
    \end{matrix}\right) \\\nonumber
    &+\left(\begin{matrix}
               \mathcal{L}^T\mathcal{L} - \id + C^{-1}& -\mathcal{L}^T + C^{-1}\mathcal{L}^T\\
         -\mathcal{L}+\mathcal{L}C^{-1}& \mathcal{L}C^{-1}\mathcal{L}^T
    \end{matrix}\right).
\end{align}
A great part of the terms cancel and the projection simplifies to:
\begin{align}
    \mathcal{P} =
    \left(\begin{matrix}
              C^{-1}& C^{-1}\mathcal{L}^T\\
         \mathcal{L}C^{-1}& \mathcal{L}C^{-1}\mathcal{L}^T
    \end{matrix}\right),
\end{align}
which may finally be rewritten as
\begin{align}
    \label{eq:major_projection_final}
    \mathcal{P} =
    \left(\begin{matrix}
        \id\\
        \mathcal{L}
    \end{matrix}\right)
    C^{-1}
    \left(\begin{matrix}
        \id &\, \mathcal{L}^T
    \end{matrix}\right).
\end{align}
Still, the evaluation of $\mathcal{P}$ might seem complicated. However, for many practical \nrepresentability{} constraints $M_i\geq 0$ the corresponding linear functional $L_i$ has a very simple form and thus also $\mathcal{P}$ becomes straightforward to evaluate. This is especially the case if one requires contraction consistency along with other affine constraints, as then only the dependence of $M_i$ on the contraction-free part of $D_{1\ldots p}$ is important and any contribution of its traces can be ignored. In Sec.~\ref{sec:purification} we explained how to implement the $D$- and the $Q$-condition. Eq.~\eqref{eq:major_projection_final} becomes in this case simply 
\begin{equation}
\label{eq:major_projection_easy_case}
\mathcal{P} = \frac{1}{2}\left(
\begin{matrix}\id&(-1)^p\id\\(-1)^p\id&\id\end{matrix}\right).
\end{equation}
Summarizing, one iteration of the purification scheme up to this point involves computing
\begin{equation}
\label{eq:prototype_iteration}
\mathcal{M}^{(k+1)} = \mathcal{M}^{(k)}-(\mathcal{P}\circ \pi_{\textnormal{def}})\mathcal{M}^{(k)}.
\end{equation}
Since $\mathcal{M}_{\textnormal{def}}$ resulted from a rejection from a convex set and $\mathcal{P}$ is the projection onto a convex set, this is an instance of the \specialterm{alternating projections algorithm}, known to converge to a point in the intersection of the corresponding convex sets \cite{Gubin1967, bauschke_alternating_projections_1993}.
There are many ways to control the algorithm's rate of convergence. For our purposes we include an additional steering factor $\alpha>0$, such that Eq.~\eqref{eq:prototype_iteration} becomes:
\begin{equation}
\label{eq:major_iteration}
\mathcal{M}^{(k+1)} = \mathcal{M}^{(k)}-\alpha (\mathcal{P}\circ \pi_{\textnormal{def}})\mathcal{M}^{(k)}.
\end{equation}
We are now ready to include the constraints that the expectation values of specific observables have to be preserved. Here, we can proceed in complete analogy to Sec.~\ref{sec:purification}.
Since the observables $X_1,\ldots, X_n$ are typically given in terms of the RDM, we have to implement these additional constraints only for the zero'th row of Eq.~\eqref{eq:major_iteration}, which reads:
\begin{equation}
\label{eq:minor_iteration}
\ppartobj^{(k+1)} = \ppartobj^{(k)}-\delta\ppartobj^{(k)},
\end{equation}
with
\begin{equation}
\label{eq:subtrahend}
\delta\ppartobj^{(k)} = \alpha (P_0\circ \pi_{\textnormal{def}})\mathcal{M}[\ppartobj^{(k)}],
\end{equation}
where
\begin{equation}
    P_0 = 
    C^{-1}
    \left(\begin{matrix}
        \id &\, \mathcal{L}^T
    \end{matrix}\right),
\end{equation}
see Eq.~\eqref{eq:major_projection_final}.
For the sake of clarity, we want to emphasize that
\begin{equation}
    \mathcal{L}^T = 
    \left(\begin{matrix}
        L_1^T &\ldots& L_n^T
    \end{matrix}\right)
\end{equation}
consists of the transposes (adjoints) of the linear maps $L_i$, where the transpose is to be taken with respect to the Hilbert-Schmidt inner product.\\
On the level of RDMs, physical observables correspond to linear functionals. By the representation theorem of Riesz, we may as well consider the linear functionals as actual vectors in the Hilbert space $\complex^{r^p\times r^p}$ of RDMs.
The statement that the purification scheme should conserve the expectation value of some operator $X$ becomes then the requirement that the subtrahend [Eq.~\eqref{eq:subtrahend}] be orthogonal to $X$. If we define the space of conserved quantities by 
\begin{equation}
\mathcal{X} := \Span{\left\{X_i \lvert i=1,\ldots, n\right\}},
\end{equation}
then all we need to do is to subtract from $\delta \ppartobj^{(k)}$ its component within $\mathcal{X}$.
We may therefore determine any orthonormal basis of $\mathcal{X}$ via the Gram-Schmidt orthogonalization procedure and use this to project $\delta \ppartobj^{(k)}$ onto the orthogonal complement of $\mathcal{X}$.
If we denote by $Y_1,\ldots,Y_n$ the orthonormal basis vectors of $\mathcal{X}$, then the new iteration update may be written as:
\begin{equation}
\label{eq:subtrahend_with_conservation}
\delta\ppartobj^{(k)} = \alpha(\Pi^\perp_\mathcal{X}\circ P_0\circ\pi_{\textnormal{def}})\mathcal{M}[\ppartobj^{(k)}],
\end{equation}
where
\begin{equation}
\label{eq:op_conserving_projection}
\Pi^\perp_\mathcal{X}(\delta\ppartobj{}) := \delta\ppartobj{} - \sum_{i=0}^n \Tr(Y_i \delta\ppartobj{}) Y_i.
\end{equation}
Using the update from Eq.~\eqref{eq:subtrahend_with_conservation} instead of Eq.~\eqref{eq:subtrahend} in Eq.~\eqref{eq:minor_iteration} amounts to one iteration of the purification algorithm. By repeating the routine several times, a purified \prdm{p} may be generated. As long as the constraints related to conserved observables are not in contradiction with the \nrepresentability{} constraints, this algorithm, being an instance of the more general alternating-projections algorithm, is guaranteed to converge.\\
Note that so far we did not mention the contraction-consistency constraint Eq.~\eqref{eq:contraction_consistency} and its implementation here. It turns out that ensuring contraction-consistency is equivalent to the conservation of all possible ($p-1$)-particle operators. As such, it is already covered by this description as one may simply add them to the set $\mathcal{X}$. The unitary decomposition is then nothing but a shortcut, resulting immediately in the orthonormalized basis, which would otherwise have to be computed with the help of Gram-Schmidt.
\section{Purification iteration formulas for the Fermi-Hubbard model}\label{ap:iteration}
In this section, we explicitly give the formulae for one step in the iterative purification scheme for the Fermi-Hubbard model within the TD2RDM from Sec.~\ref{sec:testcase}.
By inserting the expressions for the operators $Y_1$ and $Y_2$ [Eqs.~\eqref{eq:index_formula_for_y1}  and \eqref{eq:index_formula_for_y2}] into Eq.~\eqref{eq:yconserving_projection}, we arrive at explicit expressions for the subtrahend $\delta D^{(k)}_{\textnormal{def};DQ;K;\mathcal{X}}$ of the iteration formula Eq.~\eqref{eq:M_iter}. As both the interaction $W$ and $\eta^+\eta^-$, and hence also $Y_1$ and $Y_2$ depend only on the singlet component of $D_{12}$ and $Q_{12}$, the expression for the triplet component simply reads
\begin{equation}
    \sumofdandq^T_{1\ldots p;\textnormal{def};K;\mathcal{X}} = \sumofdandq^T_{1\ldots p;\textnormal{def};K}.
\end{equation}
For the singlet component, the additional conservation laws lead to the following expressions in index notation:
\begin{alignat}{2}
\forall & i:& \\\nonumber
&\left[\sumofdandq^S_{12;\textnormal{def};K;\mathcal{X}}\right]^{ii}_{ii} &&=  \left[ \sumofdandq^S_{12;\textnormal{def};K} \right] ^{ii}_{ii} - \frac{\sum_l \left[\sumofdandq^S_{12;\textnormal{def};K}\right]^{ll}_{ll}}{M_s},
\end{alignat}

\begin{alignat*}{2}
\forall & i\neq j:& \\\nonumber
&\left[\sumofdandq^S_{12;\textnormal{def};K;\mathcal{X}}\right]^{ij}_{ij} &&= \left[ \sumofdandq^S_{12;\textnormal{def};K} \right] ^{ij}_{ij} + \frac{\sum_l \left[\sumofdandq^S_{12;\textnormal{def};K}\right]^{ll}_{ll}}{M_s(M_s-1)}, \\\nonumber
&\left[\sumofdandq^S_{12;\textnormal{def};K;\mathcal{X}}\right]^{ij}_{ji} &&=  \left[ \sumofdandq^S_{12;\textnormal{def};K} \right] ^{ij}_{ji} + \frac{\sum_l \left[\sumofdandq^S_{12;\textnormal{def};K} \right]^{ll}_{ll}}{M_s(M_s-1)}, \\\nonumber
&\left[\sumofdandq^S_{12;\textnormal{def};K;\mathcal{X}}\right]^{ii}_{jj} &&= \vphantom{\frac{\sum_{k\neq l}(-1)^{k+l}\left[ \sumofdandq^S_{12;\textnormal{def};K} \right]^{kk}_{ll}}{M_s(M_s-1)}} \left[ \sumofdandq^S_{12;\textnormal{def};K} \right] ^{ii}_{jj}  - \\ \nonumber
& && (-1)^{i+j}\frac{\sum_{k\neq l}(-1)^{k+l}\left[ \sumofdandq^S_{12;\textnormal{def};K} \right]^{kk}_{ll}}{M_s(M_s-1)}.
\end{alignat*}
For any other indices $i,j,k,l$ we have:
\begin{equation}
[\sumofdandq^S_{12;\textnormal{def};K;\mathcal{X}}]^{ij}_{kl} =[\sumofdandq^S_{12;\textnormal{def};K}]^{ij}_{kl}.
\end{equation}
\section{Numerical details}\label{ap:num_details}
The purification of the \prdm{2} is applied after global time steps of $dt=0.01J^{-1}$. The iteration is stopped as soon as the defect $d$ is reduced to zero or if the maximum number of iterations of $k_\text{max}=100$ is reached. The time evolution within the global time steps is performed by the application of a time adaptive Runge-Kutta-Fehlberg propagator of fourth and fifth order to the equations of motions [Eq.~\eqref{eq:eom}]. The steering factor $\alpha$ was set to a constant value of two, increasing convergence speed. However, no systematic investigation concerning the optimal scaling factor was made. It should be noted that more advanced schemes using an adaptive $\alpha$ could further improve convergence behavior.

\bibliography{bibliography}

\end{document}